\documentclass[conference]{IEEEtran}
\IEEEoverridecommandlockouts
\usepackage{cite}
\usepackage{amsmath,amssymb,amsfonts}
\usepackage{algorithmic}
\usepackage{graphicx}
\usepackage{textcomp}
\usepackage{xcolor}
\usepackage{caption}
\usepackage{placeins}
\usepackage{subcaption}
\usepackage{float}
\usepackage{url}
\def\BibTeX{{\rm B\kern-.05em{\sc i\kern-.025em b}\kern-.08em
    T\kern-.1667em\lower.7ex\hbox{E}\kern-.125emX}}
\begin{document}

\title{Evaluating Four FPGA-accelerated Space Use Cases based on Neural Network Algorithms for On-board Inference
}

\author{\IEEEauthorblockN{1\textsuperscript{st} Pedro Antunes}
\IEEEauthorblockA{\textit{Department of Computer Science} \\
\textit{Kungliga Tekniska högskolan (KTH)}\\
Stockholm, Sweden \\
pedroa@kth.se}
\and
\IEEEauthorblockN{2\textsuperscript{nd} Muhammad Ihsan Al Hafiz}
\IEEEauthorblockA{\textit{Department of Computer Science} \\
\textit{Kungliga Tekniska högskolan (KTH)}\\
Stockholm, Sweden \\
miahafiz@kth.se}
\and
\IEEEauthorblockN{3\textsuperscript{rd} Jonah Ekelund}
\IEEEauthorblockA{\textit{Department of Computer Science} \\
\textit{Kungliga Tekniska högskolan (KTH)}\\
Stockholm, Sweden \\
jonahek@kth.se}
\and
\IEEEauthorblockN{4\textsuperscript{th} Ekaterina Dineva}
\IEEEauthorblockA{\textit{Plasma-astrophysics} \\
\textit{KU Leuven}\\
Leuven, Belgium \\
ekaterina.dineva@kuleuven.be}
\and
\IEEEauthorblockN{5\textsuperscript{th} George Miloshevich}
\IEEEauthorblockA{\textit{Plasma-astrophysics} \\
\textit{KU Leuven}\\
Leuven, Belgium \\
george.miloshevich@kuleuven.be}
\and
\IEEEauthorblockN{6\textsuperscript{th} Panagiotis Gonidakis}
\IEEEauthorblockA{\textit{Plasma-astrophysics} \\
\textit{KU Leuven}\\
Leuven, Belgium \\
panagiotis.gonidakis@kuleuven.be}
\and
\IEEEauthorblockN{7\textsuperscript{th} Artur Podobas}
\IEEEauthorblockA{\centerline{\textit{Department of Computer Science}} \\
\textit{Kungliga Tekniska högskolan (KTH)}\\
Stockholm, Sweden \\
podobas@kth.se}
}

\maketitle

\begin{abstract}
Space missions increasingly deploy high-fidelity sensors that produce data volumes exceeding onboard buffering and downlink capacity. This work evaluates FPGA acceleration of neural networks (NNs) across four space use cases on the AMD ZCU104 board. We use Vitis AI (AMD DPU) and Vitis HLS to implement inference, quantify throughput and energy, and expose toolchain and architectural constraints relevant to deployment. Vitis AI achieves up to 34.16$\times$ higher inference rate than the embedded ARM CPU baseline, while custom HLS designs reach up to 5.4$\times$ speedup and add support for operators (e.g., sigmoids, 3D layers) absent in the DPU. For these implementations, measured MPSoC inference power spans 1.5-6.75 W, reducing energy per inference versus CPU execution in all use cases. These results show that NN FPGA acceleration can enable onboard filtering, compression, and event detection, easing downlink pressure in future missions.
\end{abstract}

\section{Introduction}
\label{sec:Introduction}
Future space research missions, in both near-Earth and deep-space regimes, will leverage artificial intelligence (AI) and high-performance computing (HPC) hardware~\cite{powell2022nasa} to perform \textit{in situ} analysis of the large data volumes produced by on-board high-fidelity sensors. Processing data on the spacecraft enables transmission of only distilled results over bandwidth-constrained links to Earth~\cite{9210567}.

However, modern Neural Networks (NNs) are compute-intensive and often exceed the performance or energy envelopes of radiation-hardened general-purpose processors (CPUs)~\cite{9939566}. This gap is sharper in space, where platforms face strict power, thermal, and radiation constraints. Although radiation-hardened CPUs are common, they lack the efficiency needed for sustained NN inference. Reprogrammable logic in radiation-hardened Field-Programmable Gate Arrays (FPGAs) allows the instantiation of specialized accelerators matched to these workloads. A recent survey~\cite{my_survey} underscores the advantages of this hardware for space-borne NN deployment.

This work implements and accelerates NN algorithms on FPGAs across four distinct space-mission use cases using two complementary toolchains: Vitis AI~\cite{VitisAI} with the AMD Deep Learning Processor Unit (DPU)~\cite{xilinx2023dpuczdx8g}, and Vitis High-Level Synthesis (HLS)~\cite{vitishls}. We lower energy per inference ($E = P \times t$) by (i) reducing execution time through architectural parallelism and (ii) constraining power via smaller resource footprints and selective on-chip weight residency. These characteristics define the design space tradeoff explored in the remainder of the paper. Our main contributions are:
\begin{enumerate}
    \item FPGA accelerators for four previously unaccelerated neural-network workloads relevant to upcoming missions. These NNs include two in-situ and two remote sensing applications:
    \begin{itemize}
        \item \textbf{VAE Encoder} - Probabilistic convolutional encoder for solar vector-magnetogram (SHARP) tiles (e.g., 128$\times$256 inputs) compressing to a 6-element latent vector (1:16{,}384 ratio).
        \item \textbf{CNetPlusScalar~\cite{pynets}} - CNN plus scalar-context regressor ingesting multi-modal solar imagery (HMI + AIA 193~\AA) and recent background flux to forecast future soft X-ray flux.
        \item \textbf{ESPERTA / multi-ESPERTA~\cite{esperta}} - Six parallel lightweight models for early Solar Energetic Particle (SEP) event prediction.
        \item \textbf{MMS Neural Networks (BaselineNet, ReducedNet, LogisticNet)~\cite{ekelund2024ai}} - 3D convolution plus fully connected architectures processing 32$\times$16$\times$32 ion energy distributions (FPI instrument) to classify dayside plasma regions (SW, IF, MSH, MSP).
    \end{itemize}
    \item Empirical quantification of latency, throughput, FPGA resource utilization, and energy per inference versus CPU, showing speedups up to 34.16$\times$ and inference power as low as 1.5 W.
    \item Comparative analysis of DPU (high efficiency, restricted operator set, INT8 quantization) versus HLS (flexibility: 3D layers, sigmoid, comparators, IEEE-754 precision; lower baseline parallelism) for future space deployment.
\end{enumerate}


\section{Background}
\label{sec:Section_2}

\subsection{FPGA platforms}
FPGAs are semiconductor devices that can be reconfigured to perform specific tasks. They consist of programmable logic blocks and interconnects, allowing users to implement custom hardware designs. FPGAs are suitable for space applications due to their parallel processing capabilities and adaptability to evolving mission requirements. 

This study leverages the AMD ZCU104 development board. This board is a mid-tie`red FPGA development board specifically designed for embedded design and systems, targeting applications that require flexibility and performance. This board features a comprehensive set of peripherals (USBs, HDMIs, JTAGs, etc.) and an AMD ZU7EV MPSoC device, which comprises two distinct components: the Processing System (PS), referred to as the CPU, and the Programmable Logic (PL), denoted as FPGA.

The CPU comprises two hard processing systems (HPS): a high-performance ARM Cortex A53 and a real-time focused ARM Cortex R5. In contrast, the FPGA comprises reconfigurable logic, encompassing 504k logic cells, 38 Mb (~4.75 MB) of on-chip SRAM, and 1728 DSP slices for computations.

The ZCU104 is an ideal platform for prototyping space applications, having been extensively utilized in previous research on hardware accelerators for space~\cite{10741761,10757308,9939585,10035150}. The reconfigurable area of the ZCU104 is comparable to that of radiation-tolerant or hard FPGAs (e.g., Kintex UltraScale XQR), enabling seamless transfer of results between these devices with minimal effort. 

\subsection{Implementation Approaches}
\subsubsection{Vitis AI / AMD Deep Learning Processor Unit}
Vitis AI is a tool developed by AMD/Xilinx designed to accelerate NN algorithms on AMD/Xilinx FPGAs. It streamlines the deployment of NN models on heterogeneous hardware through a unified workflow that includes model optimization, quantization, compilation, and deployment. Vitis AI supports popular deep learning frameworks such as TensorFlow and PyTorch, allowing developers to utilize pre-trained NN models or customize and train their own. These models can be quantized using either post-training quantization (PTQ) or quantization-aware training (QAT). The PTQ method involves converting weights and activations to 8-bit integers directly, while QAT fine-tunes the model with training data to enable inference with 8-bit weight and activation values.

A critical component of Vitis AI is the DPU, a configurable IP core dedicated to convolutional neural networks (CNNs). This DPU is configured by uploading instructions equivalent to the NN model intended for execution, following compilation with Vitis AI. The DPU is a scalable and energy-efficient accelerator that integrates into AMD/Xilinx FPGAs. It achieves high-performance inference by parallelizing operations such as matrix multiplications and activations, thus reducing latency and power consumption compared to CPUs. Developers can tailor the DPU's architecture to balance resource utilization and throughput, optimizing it for specific applications like real-time processing or environments with limited resources.

\subsubsection{Vitis HLS}
AMD/Xilinx Vitis HLS is a tool that enables users to define circuit functionality through C/C++, which is then transformed by the Vitis HLS compiler into a register transfer level (RTL) hardware description. This RTL description is subsequently synthesized and mapped onto an FPGA. The primary objectives of such HLS tools are to enhance productivity and portability by automating tasks traditionally performed by hardware engineers. Crucially, these tools democratize FPGA utilization for non-hardware engineers by allowing them to describe accelerators using more abstract languages than traditional hardware description languages.

Vitis HLS supports various forms of parallelism tailored to different application scenarios. These include data-level parallelism, which is analogous to vector systems in CPUs; process-level concurrency; and streams or channels between tasks. The programmer can control the degree of parallelism through specific compiler directives, thereby guiding the compiler to produce the desired hardware configuration.

\subsection{Space Use-Cases}
\subsubsection{Description of VAE Neural Network}
\begin{figure}
    \centering
    \includegraphics[width=\linewidth]{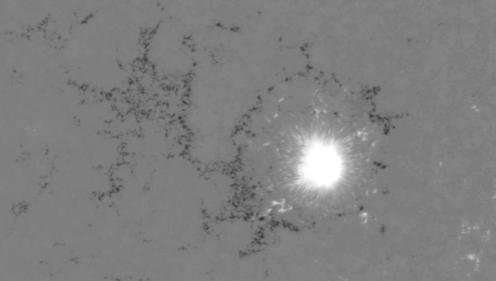}
    \caption{Example input to the VAE model: a cropped image of a solar active region showing the radial component of the Sun’s magnetic field. White areas indicate outward (positive) magnetic flux, and black areas indicate inward (negative) flux. The bright region near the center corresponds to a sunspot with strong positive polarity, surrounded by weaker areas of opposite polarity.}
    \label{fig:vae_input}
\end{figure}

The Variational Autoencoder (VAE) represents a probabilistic encoder-decoder architecture tailored for processing solar imagery. The VAE encoder transforms input images into a compact latent space, thereby reducing dimensionality while preserving essential features, which enables the decoder to reconstruct the original image or generate novel variations. This latent representation facilitates the generation of a generalized depiction of the input image, distinct from the reconstructed outputs produced by the decoder. In this context, the primary objective of employing the VAE is to project solar data, specifically from the Helioseismic and Magnetic Imager (HMI)~\cite{schou2012design} on NASA's Solar Dynamics Observatory (SDO)~\cite{SolarDynamicsObservatory}, into a latent space for visualization and analysis. This approach aids in identifying eruption precursors, signatures within active region evolution that indicate impending eruptions, and enhances the prediction/forecasting of space weather phenomena, elucidating their dynamic evolution. The VAE encoder effectively reduces the dimensionality of input data from 128x256 RGB images to a six-element array, achieving a compression ratio of 1:16,384. An example input is illustrated in Figure~\ref{fig:vae_input}.

\subsubsection{Description of CNetPlusScalar}
The CNetPlusScalar~\cite{pynets} application leverages a CNN to forecast X-ray flux from a combination of Helioseismic and Magnetic Imager (HMI) and Extreme Ultraviolet (EUV) images of the Sun, utilizing a curated machine learning dataset~\cite{Galvez_2019} crafted by Solar Dynamics Observatory (SDO) AIA and HMI instrument data. The inputs comprise 193 Å (representing coronal structures) AIA wavelengths, which undergo processing to synchronize time matching between SDO images~\cite{Galvez_2019} and GOES X-ray fluxes. A limb-brightening correction employing a geometrical function is applied to the relevant AIA wavelengths to mitigate potential biases in predictions. It frames the problem as a regression task utilizing Mean Square Error and evaluates the benefits of introducing soft constraints to predict extreme X-ray flux values more accurately. To enhance prediction capabilities, the background from preceding 30 minutes is concatenated to a fully connected layer following the convolutional layers of the CNN.

\subsubsection{Description of ESPERTA}
The ESPERTA model (Alberti et al.~\cite{esperta}), has been adapted for onboard spacecraft to predict Solar Energetic Particle (SEP) events in the inner heliosphere. The ESPERTA forecast system predicts SEP events 10 minutes after the peak of M2-class or larger solar flares using three variables: flare heliolongitude, time-integrated soft X-ray flux, and time-integrated 1 MHz radio flux. ESPERTA demonstrated a high Probability of Detection (POD) of 83\%, correctly predicting all significant SEP events, thereby confirming its effectiveness for SEP forecasting beyond Earth's orbit.

\subsubsection{Description of MMS Neural Networks}
The Magnetospheric Multiscale Mission Neural Networks (MMS NNs) proposed by Ekelund et al.~\cite{ekelund2024ai} comprise three network topologies: Baseline, Reduced, and Logistic. The Baseline topology was originally presented by Olshevsky et al.~\cite{Olshevsky_2021}; the Reduced and Logistic networks improve on the Baseline network by reducing the number of parameters by $>$ 95\% while retaining the same accuracy. The Baseline and Reduced networks are CNNs with 3D convolutional layers followed by fully connected layers. The input for the networks is a 32$\times$16$\times$32 matrix with the 3D ion energy distribution obtained by the Fast Plasma Investigation instrument (FPI)~\cite{pollockfastplasmainvestigation2016} onboard the MMS spacecraft. The networks classify the input data as belonging to one of four regions from the Earth's dayside plasma environment: SW (solar wind), IF (ion foreshock), MSH (magnetosheath), or MSP (magnetosphere). The primary use cases for these networks are the selective downlink of scientific data onboard the spacecraft and identifying a region of interest (ROI) for high-rate and high-precision data collection. Throughout this work, we refer to them as BaselineNet, ReducedNet, and LogisticNet.

\section{Methodology}
\label{sec:Section_3}
The neural network algorithms accelerated on the FPGA exhibit distinct characteristics. Some are designed for in-situ applications, while others target remote sensing. Nevertheless, all of them have the potential to reduce onboard storage requirements and downlink demand. The ESPERTA, CNetPlusScalar, and MMS networks each output a single scalar value that enables decision-making actions, such as data discarding or selective downlinking. In contrast, the VAE encoder compresses input images into compact latent representations suitable for downlink and ground-based decoding. Among these models, ESPERTA is the simplest, CNetPlusScalar is the most complex in terms of parameters and operations, and the MMS networks lie in between. The latter introduce 3D layers that are difficult to accelerate and largely unsupported by existing space-focused accelerators.

\subsection{Neural Networks Implemented}
All networks largely retain their original topologies. Their sizes range from 24 to 3,061,966 parameters. Table~\ref{tab:nn_p_o} summarizes the number of parameters and operations for each model. We generated their graphical representations using Netron~\cite{roederNetronVisualizerNeural2025}.

\begin{table}[!ht]
    \centering
    \caption{Summary of parameters and operations.}
    \begin{tabular}{lll}
    \hline
        \textbf{} & \textbf{\# Parameters} & \textbf{\# Operations} \\ \hline
        \textbf{VAE Encoder} & 395,692 & 83,417,100 \\
        \textbf{CNetPlusScalar} & 3,061,966 & 918,241,400 \\
        \textbf{multi-ESPERTA} & 24 & 60 \\
        \textbf{LogisticNet} & 8,196 & 30,720 \\
        \textbf{ReducedNet} & 44,624 & 502,961 \\
        \textbf{BaselineNet} & 915,492 & 110,541,696 \\ \hline
    \end{tabular}
    \label{tab:nn_p_o}
\end{table}

\subsubsection{VAE Encoder}
\begin{figure*}[!ht]
    \centering
    \includegraphics[width=0.8\linewidth]{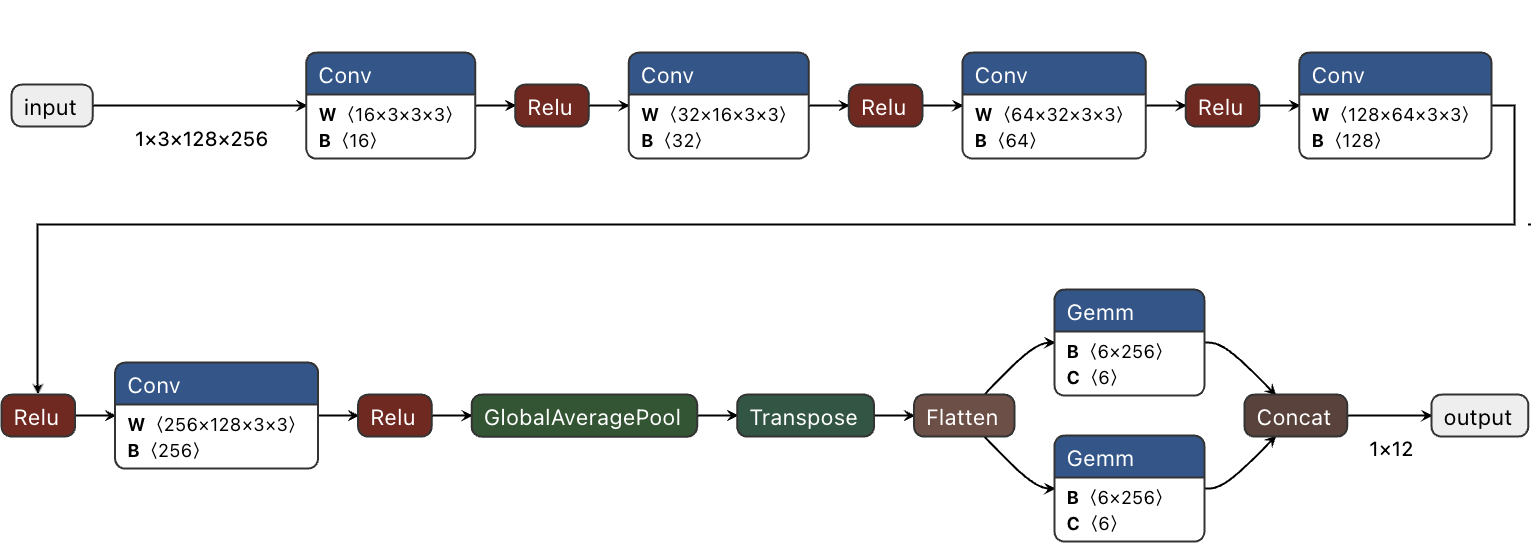}
    \caption{VAE encoder architecture (sampling and exponent handled on CPU).}
    \label{fig:vaemodel1}
\end{figure*}

We omit the final two encoder operations (random sampling and exponent) from FPGA accelerator and execute them on the CPU. These operations were unsuitable to map to FPGA. The NN architecture which we accelerated is depicted in Figure~\ref{fig:vaemodel1}.

\subsubsection{CNetPlusScalar}
\begin{figure*}[!ht]
    \centering
    \includegraphics[width=0.9\linewidth]{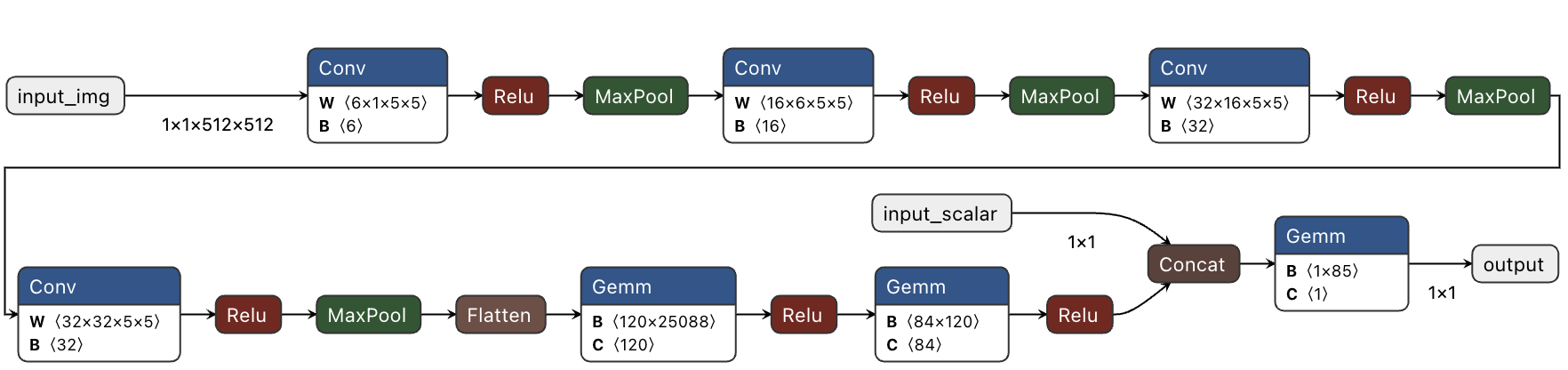}
    \caption{CNetPlusScalar architecture after replacing leaky ReLU with ReLU.}
    \label{fig:CNetPlusScalar}
\end{figure*}

We replaced Leaky ReLU layers which are unsupported by Vitis AI with standard ReLU. The modified architecture is shown in Figure~\ref{fig:CNetPlusScalar}.

\subsubsection{ESPERTA}
\begin{figure}[!ht]
    \centering
    \includegraphics[width=\linewidth]{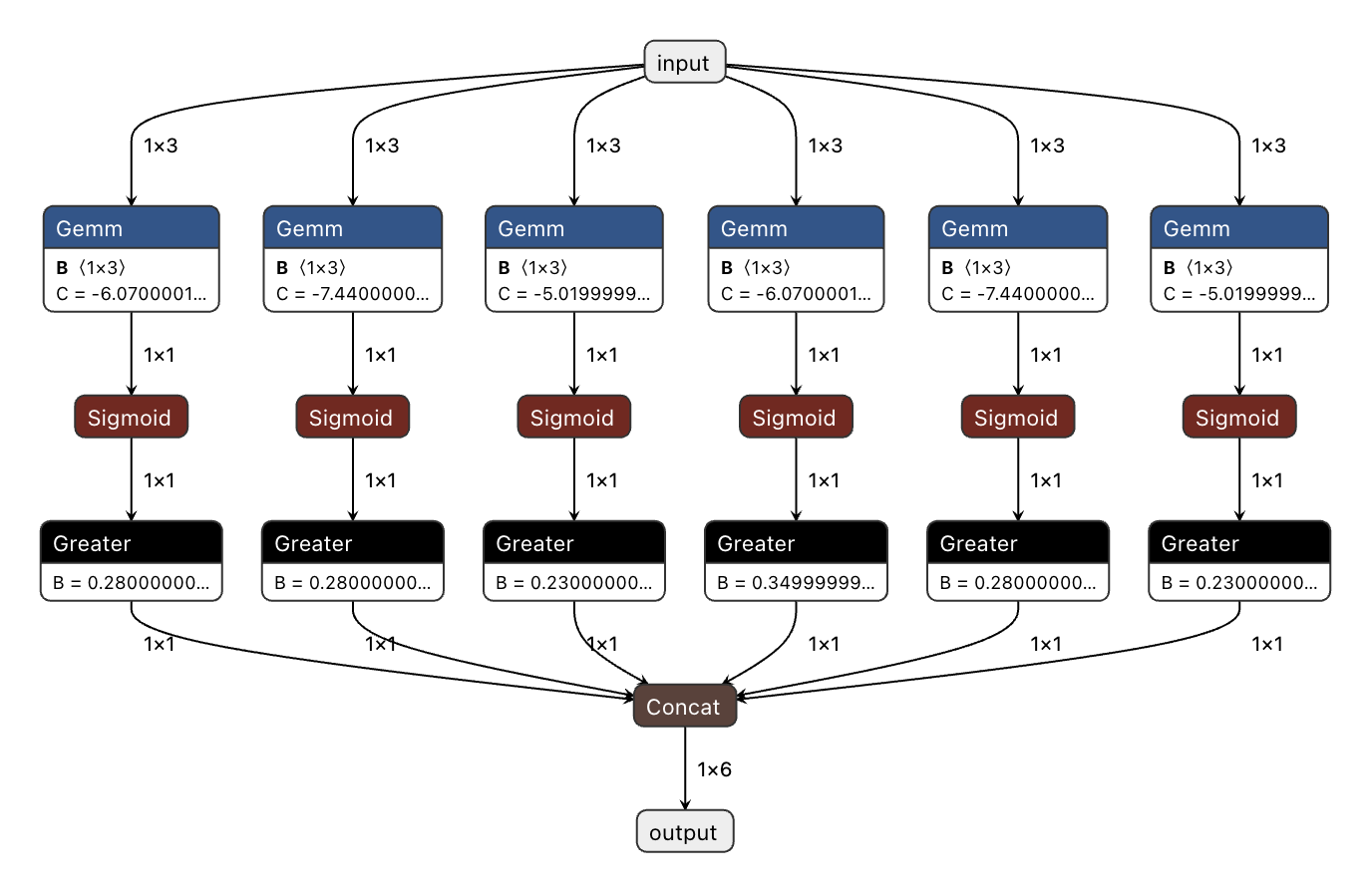}
    \caption{Parallel multi-ESPERTA architecture (six shared-input models).}
    \label{fig:Multi_ESPERTA}
\end{figure}

The original algorithm invoked six ESPERTA models with different parameters sequentially. These models were put together in parallel to form a new network, multi-ESPERTA. This parallel form reduces control overhead and suits FPGA mapping. We implemented ESPERTA and multi-ESPERTA in PyTorch and set weights and thresholds per Laurenza et al.~\cite{esperta_weights}. Figure~\ref{fig:Multi_ESPERTA} shows the topology.

\subsubsection{MMS Neural Networks (LogisticNet, ReducedNet, BaselineNet)}
\begin{figure*}[!ht]
    \centering
    \includegraphics[width=0.90\linewidth]{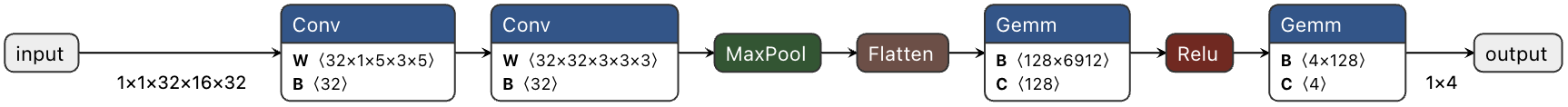}
    \caption{BaselineNet architecture.}
    \label{fig:BaselineNet}
\end{figure*}

\begin{figure*}[!ht]
    \centering
    \includegraphics[width=0.80\linewidth]{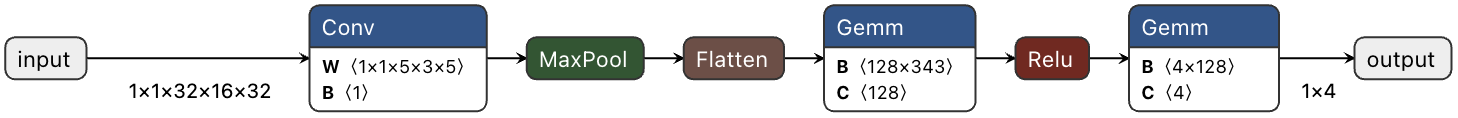}
    \caption{ReducedNet architecture.}
    \label{fig:ReducedNet}
\end{figure*}

\begin{figure*}[!ht]
    \centering
    \includegraphics[width=0.6\linewidth]{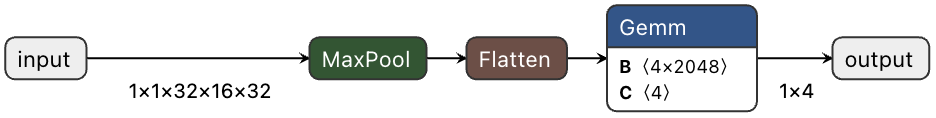}
    \caption{LogisticNet architecture.}
    \label{fig:LogisticNet}
\end{figure*}

We removed the final sigmoid since classification depends only on the argmax of logits. Figures~\ref{fig:BaselineNet}, \ref{fig:ReducedNet}, and \ref{fig:LogisticNet} show the modified architectures.


\subsection{Implementing Neural Networks on FPGAs}
The implementation strategy has a significant impact on inference performance. Vitis AI offers aggressive, domain-specific optimizations tailored for CNNs. We used Vitis AI to implement the VAE encoder and CNetPlusScalar networks. However, Vitis AI does not support ESPERTA due to its use of sigmoid and "greater-than" layers, or the MMS networks because they rely on 3D pooling and convolutional layers. For these models, we employed Vitis HLS to explore the unsupported topologies. HLS enables integration of custom C code, supports a wider range of architectures, and maintains IEEE-754 precision where quantization would otherwise reduce accuracy. Unoptimized HLS implementations, however, exploit limited parallelism.

We developed a Jupyter notebook that ran on the ZCU104 board within the AMD PYNQ environment. The notebook executed the PyTorch model on both the CPU and the FPGA. In this environment, we measured (i) inference latency, (ii) total board power through the 12 V rail (Infineon IRPS5401), and (iii) MPSoC power through the INT rail. Each measurement was performed after a cold reset. The energy per inference was computed as $E = P_{\text{MPSoC}} \times t$.

We also evaluated FINN~\cite{umuroglu2017finn} using Brevitas~\cite{brevitas} QONNX exports, but unsupported operations and failed layer conversions prevented the generation of the NN accelerators.

\subsubsection{Vitis AI}
We compile PyTorch models into XIR (.xmodel) using Vitis AI 2.5 (Docker) and load them onto a DPU IP (DPUCZDX8G B4096) built in Vivado 2022.1 via the DPU-PYNQ reference design. The build produces dpu.bit, dpu.hwh, and dpu.xclbin, which we load with the DPUOverlay API to set inputs and read outputs. In a Python notebook we load weights, set the DPU target (DPUCZDX8G\_ISA1\_B4096), run the inspector to verify that all layers are supported, and apply PTQ.

\subsubsection{HLS}
We used Vitis HLS 2022.1 to generate an IP core for each NN and Vivado 2022.1 to integrate it into the Zynq platform. We first converted each PyTorch model to ONNX, then to C code with ONNX2C~\cite{onnx2c}. We adopted a naive translation to HLS (no performance directives), adding interface pragmas to expose input tensor, output tensor, and control registers (start, auto-start, done, interrupt enable) as memory-mapped (AXI4-Lite) registers. For large inputs, we instead exposed a register holding a DRAM address shared with the CPU, letting the accelerator fetch data via an AXI4 master. By default, we instantiated all weights on-chip; weights that did not fit in BRAM were placed in DRAM. The resulting accelerator follows a streaming (dataflow) organization, mapping each layer onto the fabric.

\begin{figure}[!ht]
    \centering
    \includegraphics[width=0.48\textwidth]{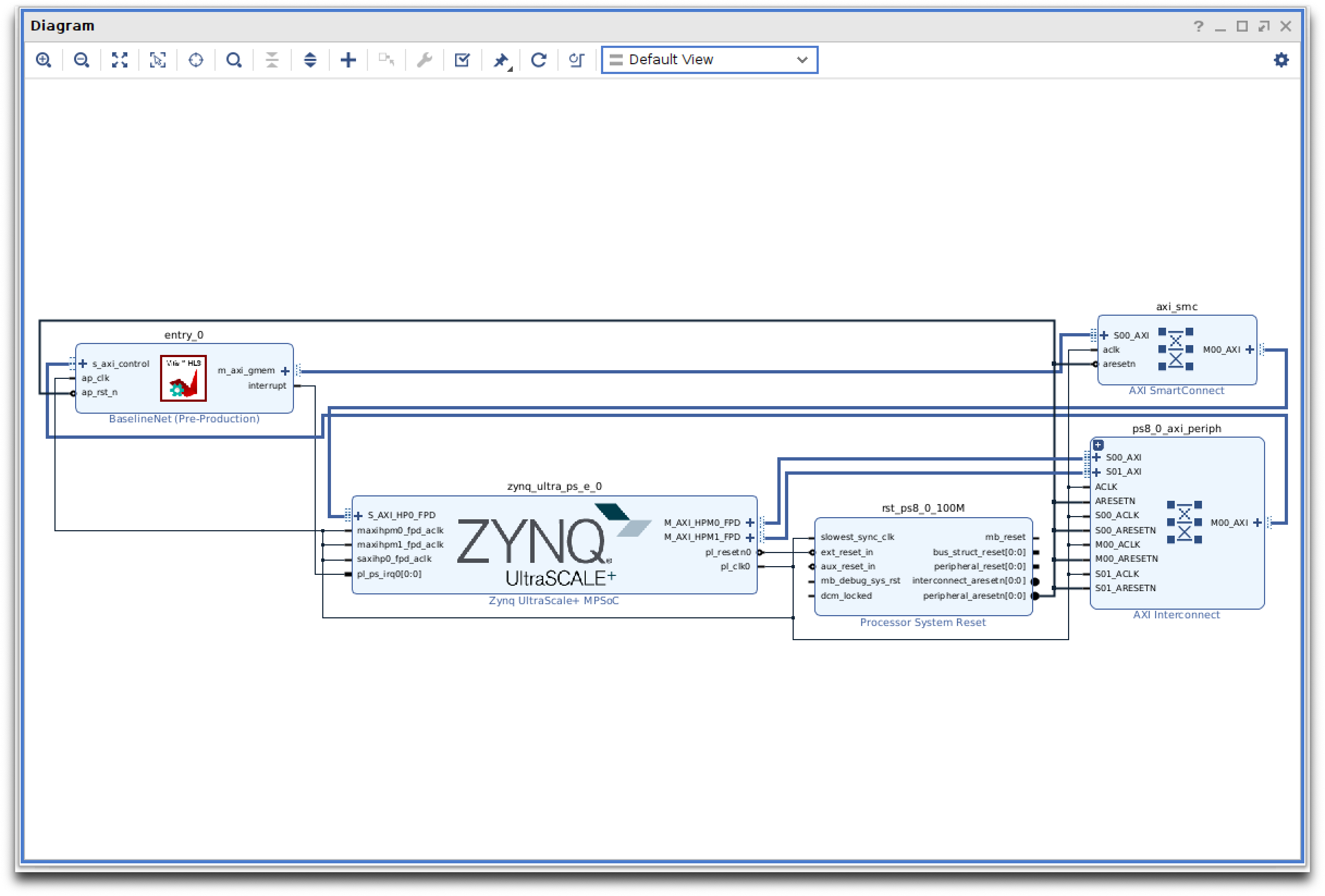}
    \caption{Block diagram of our neural network accelerators integrated with Zynq MPSoC.}
    \label{fig:BlockDiagram}
\end{figure}

We then integrated the generated IP core with the Zynq MPSoC in a Vivado block design. Vivado auto-generated the AXI interconnects linking master and slave interfaces (Figure~\ref{fig:BlockDiagram}), after which it performed synthesis, placement, routing, and bitstream generation, producing the .bit and .hwh files required to program the FPGA.

Interaction with this IP mirrors the Vitis AI DPU flow. Within a PYNQ Jupyter notebook, we load the bitstream using the Overlay class, allocate buffers for inputs and weights, and invoke three helper functions: \textit{load\_ip\_input()}, \textit{start\_ip()}, and \textit{read\_ip\_output()}. These write/read the necessary MMIO registers to stage input data, trigger execution, and fetch outputs. The \textit{load\_ip\_input()} and \textit{read\_ip\_output()} functions also handle lightweight pre- and post-processing. The \textit{start\_ip()} function asserts the start bit and polls the done bit until the hardware signals completion.

\section{Results}
\label{sec:Section_4}
Implementation results show that the Vitis AI approach delivers high performance, while the Vitis HLS approach offers greater flexibility, supporting a broader range of NN topologies and higher-precision arithmetic. These results highlight a trade-off between throughput and operator coverage/precision. This section details each implementation's resource utilization, performance metrics, and power consumption.

Table~\ref{tab:resource_consumption} reports resource utilization and shows a pronounced gap between the Vitis AI and Vitis HLS implementations. This gap reflects both the DPU's specialized architecture and the lack of explicit parallelism exploration in our HLS designs. Given this lack of explicit parallelism, the tool defaults to a safe method for mapping the C code to RTL. This method typically means executing tasks sequentially, which may lead to longer inference times. 

The low BRAM usage of ESPERTA and LogisticNet matches their small parameter counts. In our HLS implementations this resource and the NNs parameters are linked since we attempt to store all parameters on chip.  ESPERTA uses 1.5 BRAMs (one 36 Kb and one 18 Kb block). LogisticNet uses 13 BRAMs, this corresponds to 58.5 KB of on-chip memory. Given that its parameters occupy $\approx$33KB, we predict that the rest of the on-chip memory is used between layers for intermediate feature maps. This on-chip policy inflates BRAM usage for larger HLS models (e.g., ReducedNet and BaselineNet) and pushes BaselineNet to external memory because its parameters do not fit. Fetching these parameters from external memory can further increase inference time. By comparison, the AMD DPU in the Vitis AI implementations stores parameters in BRAMs and UltraRAMs. This hardware uses 165 BRAMs and 92 UltraRAMs, which is approximately 3.92 MB of on-chip memory.

Table~\ref{tab:performance_report} lists performance metrics. Every design outperforms the CPU baseline except ReducedNet and BaselineNet under HLS. Although all HLS designs run at 100 MHz, deeper networks require more cycles because the tool inserts additional stages to meet timing, which increases latency. This behavior indicates that a naive HLS mapping is suitable mainly for shallow, lightweight models. A benefit of HLS over Vitis AI is numerical fidelity: the CPU and HLS outputs for ESPERTA and the MMS networks match within $\le 10^{-10}$, which is valuable for onboard decision logic. These HLS designs preserve 32-bit floating-point precision. Vitis AI, in contrast, relies on INT8 quantization, which introduces approximation error; in our case, PTQ caused noticeable degradation that QAT could mitigate.

In contrast, the Vitis AI implementation achieves substantial speedups over the ARM CPU for larger models: $24.06\times$ for the VAE Encoder and $34.16\times$ for CNetPlusScalar. Both models map well to Vitis AI because all parameters fit on chip and all layers are supported by the DPU. We also explored minor CNetPlusScalar modifications to resemble the VAE Encoder: (i) removing the pooling layers and (ii) reducing parameter and operation counts to similar levels. Both CPU and DPU inference became faster, but the speedup shrank, indicating that the CPU benefited more from these changes than the DPU. This helps explain why the VAE Encoder's speedup is lower than CNetPlusScalar's. Finally, removing the scalar input in CNetPlusScalar had negligible effect: CPU performance was indistinguishable, and Vitis AI improved by approximately 2 frames per second (FPS).

Table~\ref{tab:performance_report} distinguishes between board and MPSoC power, showing that the MPSoC draws only a fraction of the total board power. The board’s power consumption includes the power used by peripherals that might not be present on boards used in space missions.

\begin{figure}[!t]
    \centering
    \includegraphics[width=\linewidth]{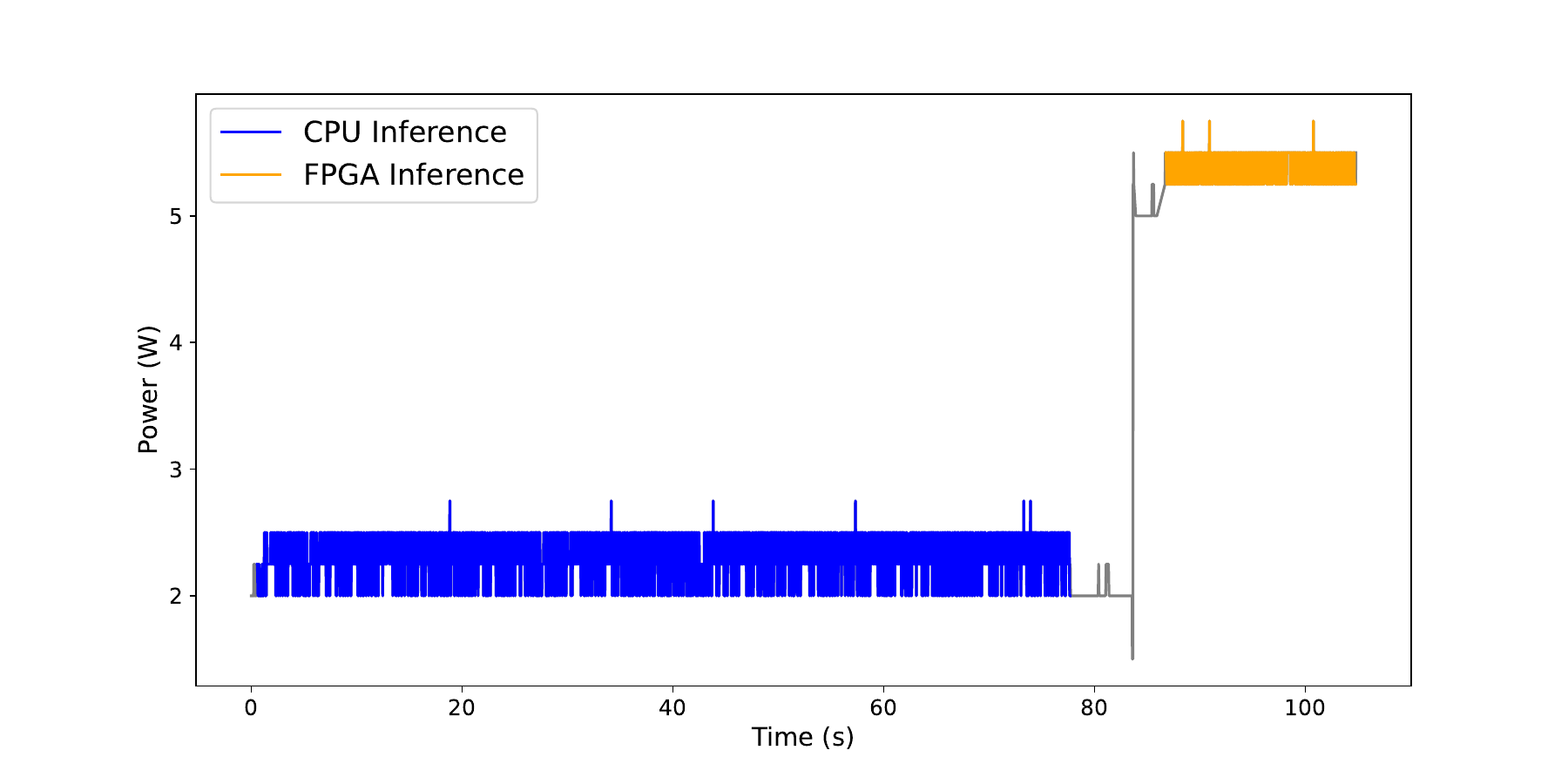}
    \caption{ZCU104 (MPSoC) power consumption during inference of the VAE Encoder (1000 inputs).}
    \label{fig:power_consumption_vaenn}
\end{figure}

\begin{figure}[!t]
    \centering
    \includegraphics[width=\linewidth]{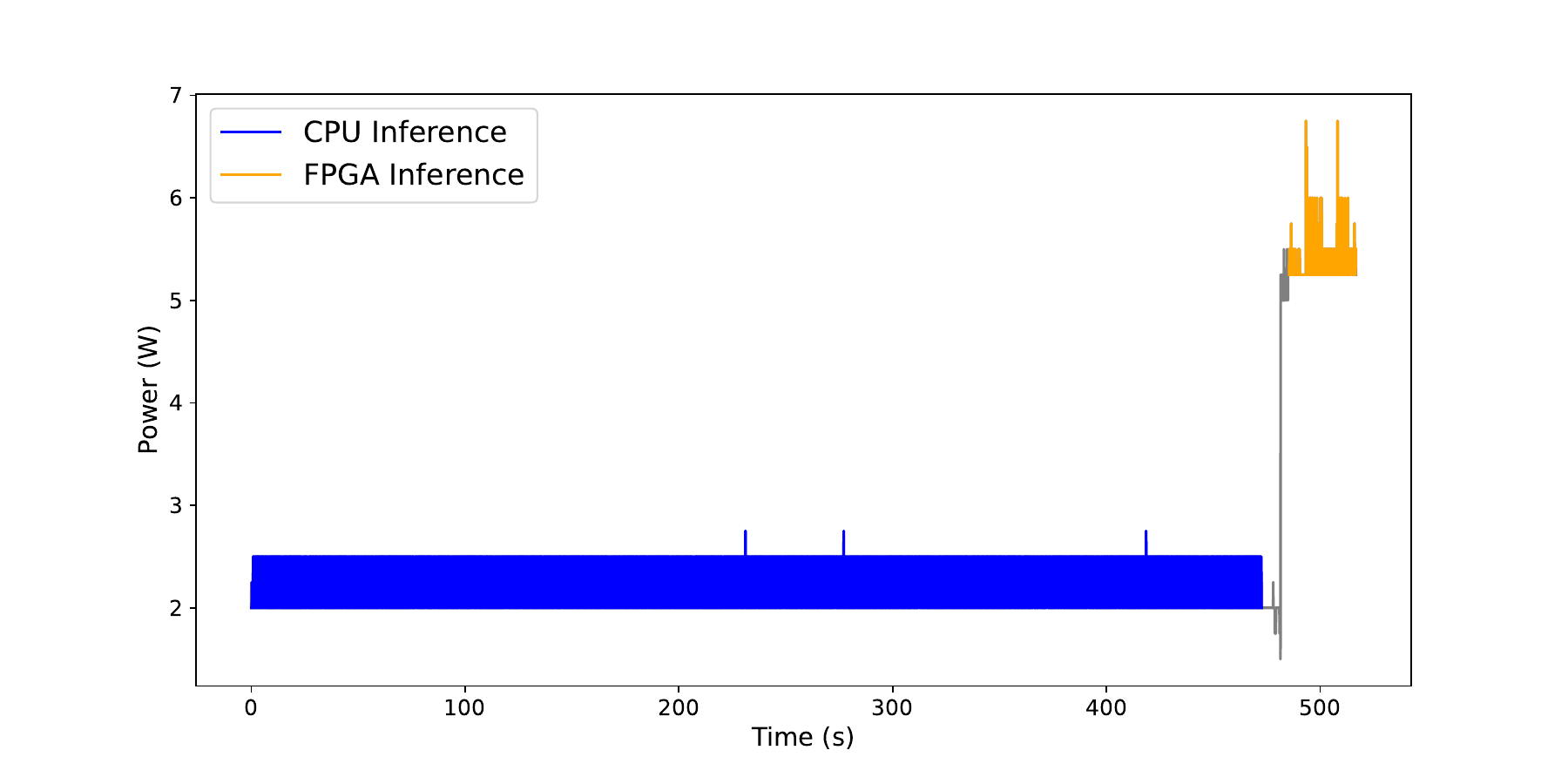}
    \caption{ZCU104 (MPSoC) power consumption during inference of the CNetPlusScalar network (1000 inputs).}
    \label{fig:power_consumption_cnet}
\end{figure}

\begin{figure}[!t]
    \centering
    \includegraphics[width=\linewidth]{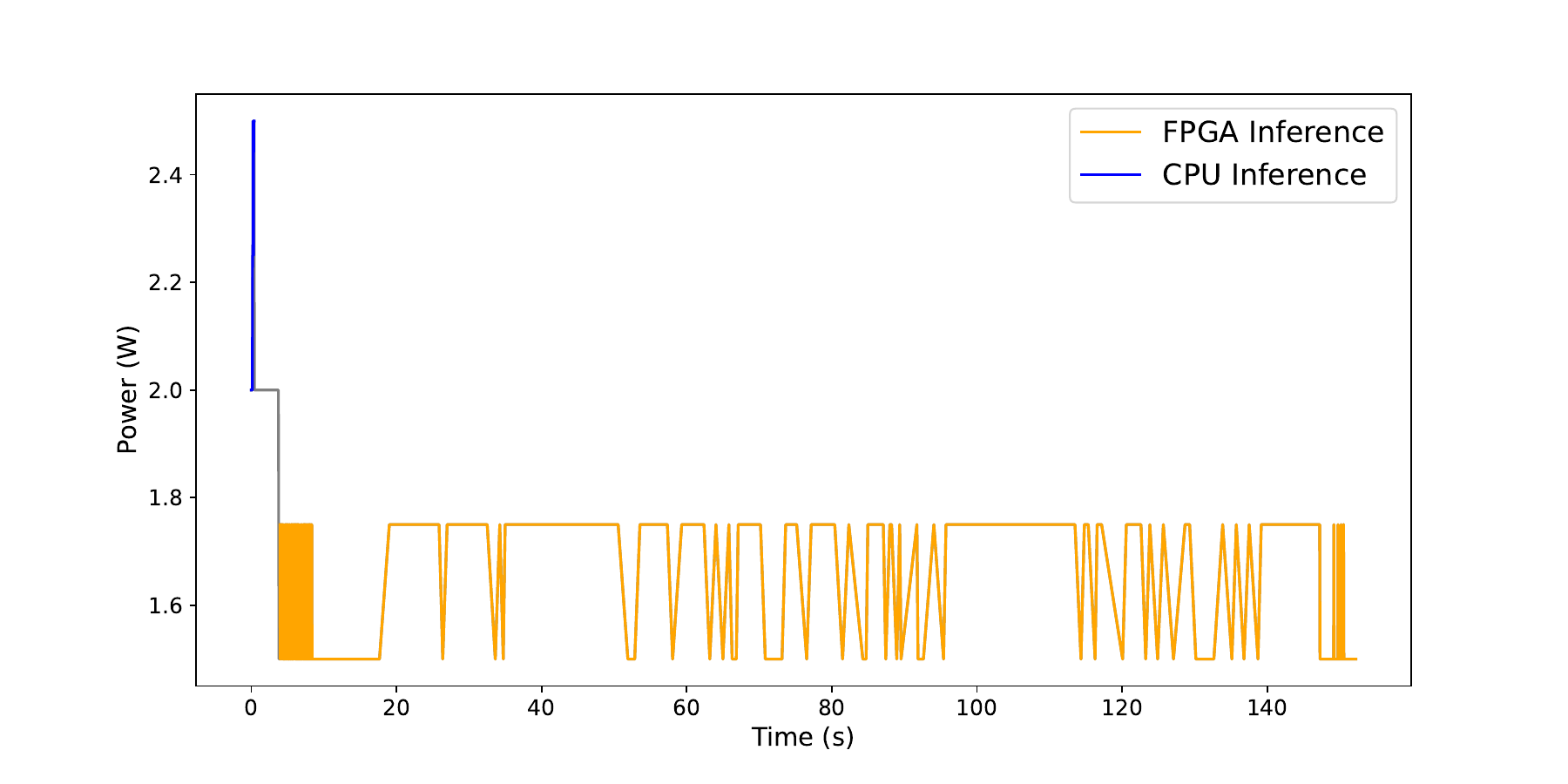}
    \caption{ZCU104 (MPSoC) power consumption during inference of the multi-ESPERTA network (1000000 inputs).}
    \label{fig:power_consumption_esperta}
\end{figure}

\begin{figure}[!t]
    \centering
    \includegraphics[width=\linewidth]{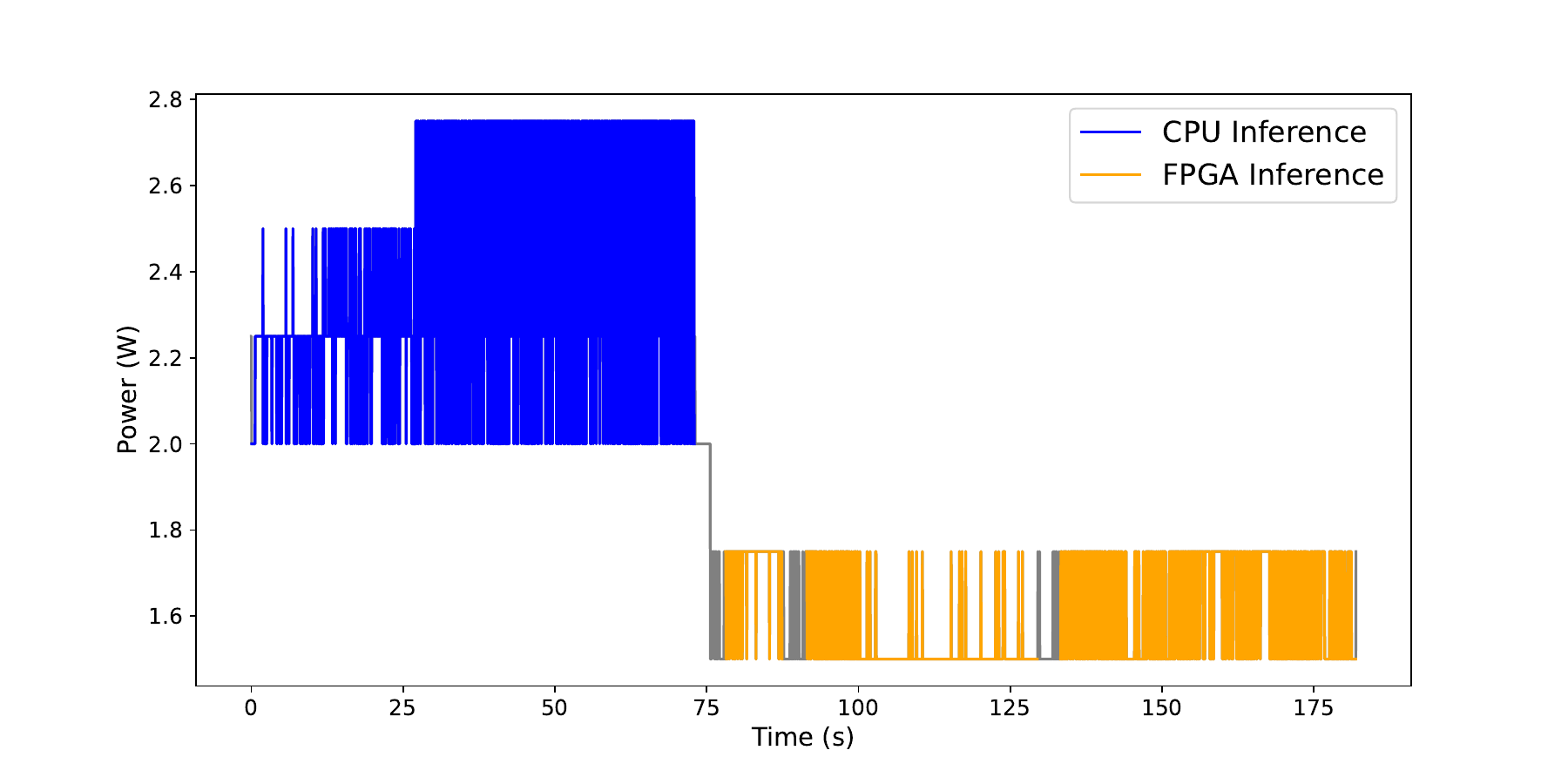}
    \caption{ZCU104 (MPSoC) power consumption during inference of MMS Neural Networks (LogisticNet, ReducedNet, BaselineNet). BaselineNet used 10 inputs due to higher computational complexity (others used 1000).}
    \label{fig:power_consumption_mmsnn}
\end{figure}

Figures~\ref{fig:power_consumption_vaenn}--\ref{fig:power_consumption_mmsnn} show power consumption over time for CPU and FPGA runs (CPU windows in blue, FPGA in orange, and other phases in grey). The CPU draws more power than the lightweight HLS accelerators but less than the DPU-based designs.

Figures~\ref{fig:power_consumption_vaenn} and~\ref{fig:power_consumption_cnet} illustrate the impact of FPGA programming (bitstream loading) on total MPSoC power consumption. In these cases, the dynamic FPGA inference power does not always exceed that of the CPU. The dynamic component can be observed in the figures as the variation between minimum and maximum power during inference. Nevertheless, adding the DPU increases static power, leading to higher overall power consumption. In contrast, the compact HLS accelerators (Figures~\ref{fig:power_consumption_esperta} and~\ref{fig:power_consumption_mmsnn}) reduce both static and dynamic power, resulting in lower inference energy than the CPU. CPU measurements are taken immediately after reboot, before any bitstream is programmed.

In Figure~\ref{fig:power_consumption_esperta}, loading the inputs to the accelerator takes more time to execute than the inference itself. When measuring the inference time, we did not consider this time since, in a real-world scenario, these inputs would come directly from the sensors to the accelerator. Nevertheless, they appear in the figure as FPGA inference, since when measuring the power consumption, we cannot distill the accelerator inference from the loading due to its low inference time.

\begin{figure}[t]
    \centering
    \includegraphics[width=\linewidth]{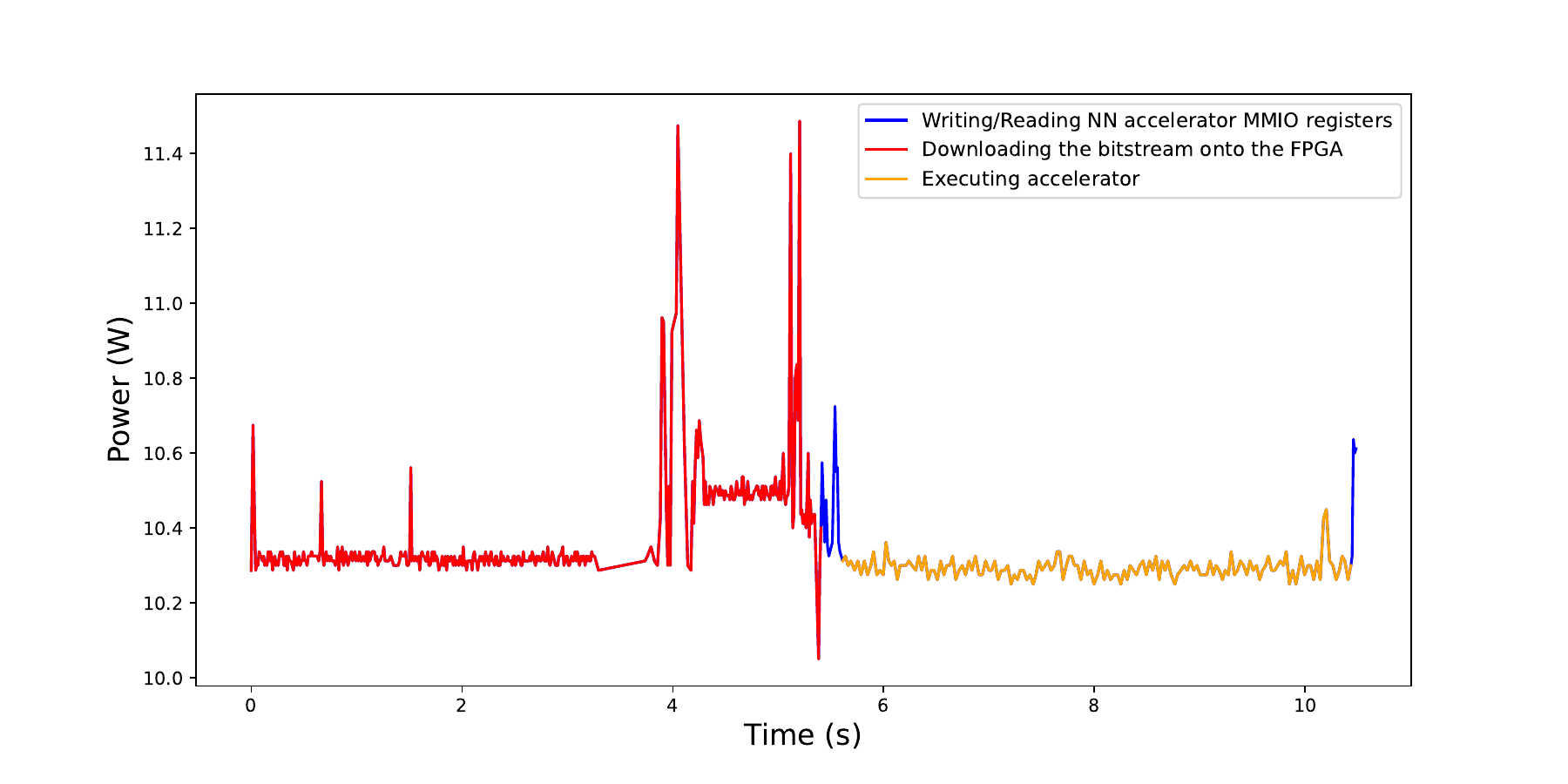}
    \caption{Detailed Power Consumption of the BaselineNet on the ZCU104 board}
    \label{fig:power_consumption_profile}
\end{figure}

Figure~\ref{fig:power_consumption_profile} decomposes the power consumption of the entire board during a single inference interval. The dominant power spike occurs during bitstream configuration, while the lowest draw appears when the CPU waits for the accelerator to complete. The peak power consumption during FPGA programming (bitstream download) is an important factor in space mission planning. This is particularly relevant when FPGA scrubbing is used to periodically reprogram the device and mitigate radiation-induced bit flips in the FPGA configuration.

\begin{table*}[t]
    \centering
    \caption{Resource Utilization and Clock Frequency of Implementations on the ZCU104 FPGA}
    \begin{tabular}{lllllll}
    \hline
        \textbf{} & \textbf{LUTs} & \textbf{FFs} & \textbf{DSPs} & \textbf{BRAMs} & \textbf{URAMs} & \textbf{Clock/s Frequency} \\ \hline
        \textbf{Available Resources} & 230,000 & 461,000 & 1,728 & 312 & 96 & - \\ 
        \textbf{B4096 DPU (Vitis AI)} & 102,154 (44\%) & 199,192 (43\%) & 1,420 (82\%) & 165 (52\%) & 92 (95\%) & 75/100/300/600 MHz \\ 
        \textbf{ESPERTA HLS} & 8,096 (4\%) & 12,052 (3\%) & 36 (2\%) & 1.5 (1\%) & - & 100 MHz \\ 
        \textbf{LogisticNet HLS} & 5,382 (2\%) & 6,476 (1\%) & 5 (0.29\%) & 13 (4\%) & - & 100 MHz \\ 
        \textbf{ReducedNet HLS} & 5,395 (2\%) & 7,190 (2\%) & 16 (1\%) & 68.5 (22\%) & - & 100 MHz \\ 
        \textbf{BaselineNet HLS} & 6,500 (3\%) & 8,243 (2\%) & 19 (1\%) & 150.5 (48\%) & - & 100 MHz \\ \hline
    \end{tabular}
    \label{tab:resource_consumption}
\end{table*}

\begin{table*}[t]
    \centering
    \caption{Performance metrics for each application implemented on the ZCU104 board}
    \begin{tabular}{lllllll}
    \hline
        \textbf{} & \textbf{Speed up} & \textbf{FPS} & \textbf{Throughput} & \textbf{P$_{Board}$} & \textbf{P$_{MPSoC}$} & \textbf{Energy per Inference} \\ \hline
        \textbf{VAE Encoder - CPU} & 1$\times$ & 25.21 & 2.103 MOP/s & 12.125 W & 2.75 W & 109.08 mJ \\ 
        \textbf{VAE Encoder - Vitis AI} & 24.06$\times$ & 606.65 & 50,604 MOP/s & 15.337 W & 5.75 W & 9.48 mJ \\ 
        \textbf{CNetPlusScalar - CPU} & 1$\times$ & 4.79 & 4,398 MOP/s & 12.862 W & 2.75 W & 574.11 mJ \\ 
        \textbf{CNetPlusScalar - Vitis AI} & 34.16$\times$ & 163.51 & 150,142 MOP/s & 15.987 W & 6.75 W & 41.28 mJ \\ 
        \textbf{ESPERTA - CPU} & 1$\times$ & 6,932 & 0.415 MOP/s & 11.725 W & 2.0 W & 0.29 mJ \\ 
        \textbf{ESPERTA - HLS} & 5.33$\times$ & 37,231 & 2 MOP/s & 10.6 W & 1.5 W & 0.04 mJ \\ 
        \textbf{LogisticNet - CPU} & 1$\times$ & 319 & 9 MOP/s & 11.725 W & 2.25 W & 7.03 mJ \\ 
        \textbf{LogisticNet - HLS} & 2.03$\times$ & 646 & 19 MOP/s & 10.7 W & 1.75 W & 2.71 mJ \\ 
        \textbf{ReducedNet - CPU} & 1$\times$ & 186 & 93 MOP/s & 11.9 W & 2.25 W & 12.05 mJ \\ 
        \textbf{ReducedNet - HLS} & 0.16$\times$ & 30 & 15 MOP/s & 10.512 W & 1.5 W & 49.73 mJ \\ 
        \textbf{BaselineNet - CPU} & 1$\times$ & 42 & 4753 MOP/s & 12.725 W & 2.75 W & 63.45 mJ \\ 
        \textbf{BaselineNet - HLS} & 0.01$\times$ & 0.21 & 23 MOP/s & 10.537 W & 1.75 W & 8467.82 mJ \\ \hline
    \end{tabular}
    \label{tab:performance_report}
\end{table*}

\section{Related Work}
\label{sec:Section_5}
Several studies have explored onboard computing with Vitis AI for CNN workloads. In this study we limit our comparison to other implementation of space use cases. Papatheofanous et al.~\cite{9939585} and Cratere et al.~\cite{10851354} segmented clouds in satellite imagery. Zhao et al.~\cite{app131810111} applied CNN-based object detection on spacecraft, while Ekblad et al.~\cite{10115681} detected satellite components. Coca and Datcu~\cite{10119157} addressed anomaly detection in remote images, and Castelino et al.~\cite{10741761} used convolutional autoencoders for hyperspectral artifact identification. In contrast, this work applies a CNN (CNetPlusScalar) for X-ray flux forecasting and a VAE encoder for eruption precursor analysis. Table~\ref{tab:comparation_other_work_vitisai} summarizes the results from these studies.

\begin{table}[t]
    \centering
    \caption{Comparison of Vitis AI performance with related work}
    \begin{tabular}{lllll}
    \hline
        \textbf{Network} & \textbf{Board} & \textbf{\# Param.} & \textbf{FPS} & \textbf{Power} \\ \hline
        \textbf{VAE Encoder} & ZCU104 & 395,692 & 606 & 5.75 W \\
        \textbf{CNetPlusScalar} & ZCU104 & 3,061,966 & 163 & 6.75 W \\
        \textbf{LD-UNet~\cite{9939585}} & ZCU104 & 5,652 & 632 & 14.1 W \\
        \textbf{CAE~\cite{10741761}} & ZCU104 & 2,950,000 & 250 & 5.3 W \\
        \textbf{ResNet-50~\cite{10119157}} & ZCU102 & - & 51-85 & 30 W \\
        \textbf{mod. YOLOv4~\cite{10115681}} & KV260 & - & 3.8 & - \\
        \textbf{YOLOv4-Mobv3~\cite{app131810111}} & KV260 & 5,690,000 & 48 & 7.2 W \\
        \textbf{Pixel-Net~\cite{10851354}} & Ultra96-V2 & 17,430 & 0.051 & 2.4 W \\
        \textbf{Patch-Net~\cite{10851354}} & Ultra96-V2 & 13,000 & 0.049 & 2.5 W \\
        \textbf{Scene-Net~\cite{10851354}} & Ultra96-V2 & 3,320,000 & 57 & 2.5 W \\
        \textbf{U-Net~\cite{10851354}} & Ultra96-V2 & 26,620 & 37 & 2.4 W \\ \hline
    \end{tabular}
    \label{tab:comparation_other_work_vitisai}
\end{table}

Several studies used HLS or FINN to optimize NN performance on FPGAs. Li et al.~\cite{10757308} accelerated a CNN for light-field depth extraction using FINN. Kim et al.~\cite{kim2024orbit} implemented TriCloudNet (TCN) and a U-Net-like model on a Zynq-7020 device using HLS for onboard cloud detection. Table~\ref{tab:comparation_other_work_hls} summarizes the results from these studies.

\begin{table}[t]
    \centering
    \caption{Comparison of HLS performance with related work}
    \begin{tabular}{lllll}
    \hline
        \textbf{Network} & \textbf{Board} & \textbf{\# Param.} & \textbf{FPS} & \textbf{Power} \\ \hline
        \textbf{multi-ESPERTA} & ZCU104 & 24 & 37,231 & 1.5 W \\
        \textbf{LogisticNet} & ZCU104 & 8,196 & 646 & 1.75 W \\
        \textbf{CNN~\cite{10757308}} & ZCU104 & 245,000 & 3,676 & 9.493 W \\
        \textbf{TCN+U-Net~\cite{kim2024orbit}} & Z-7020\footnotemark & 2,000 & 0.98 & 0.196 W \\ \hline
    \end{tabular}
    \label{tab:comparation_other_work_hls}
\end{table}
\footnotetext{Only the Zynq-7020 device was reported in the source; the specific development board was not specified.}

Direct comparison is difficult due to differing architectures and parameter counts that influence FPS. Few works in the literature report operations per second. This metric is useful to compare NN performance. Despite this, our implementations achieve competitive FPS relative to parameters and power, as shown in Tables~\ref{tab:comparation_other_work_vitisai} and~\ref{tab:comparation_other_work_hls}. Some studies report board power~\cite{9939585,10119157,app131810111}, while others report MPSoC power~\cite{10741761,10851354}. For broader context, see survey work~\cite{my_survey}.

\section{Conclusion}
\label{sec:Section_6}
This work presented a comprehensive evaluation of FPGA-based NN acceleration across four space-mission use cases using two complementary toolchains: Vitis AI (DPU-based) and Vitis HLS. Vitis AI achieved substantial speedups—24.06× for the VAE Encoder and 34.16× for CNetPlusScalar—demonstrating strong suitability for large convolutional workloads. However, it lacked support for ESPERTA and the MMS networks, which require sigmoid, comparison, and 3D convolution/pooling layers. Vitis HLS enabled these unsupported workloads, achieving 5× and 2× speedups for ESPERTA and LogisticNet, respectively, on the ZCU104, while preserving 32-bit floating-point accuracy. Both approaches reduced energy per inference relative to CPU execution, validating FPGA-based NN acceleration as a practical approach for onboard filtering, compression, forecasting, and event classification.

The ZCU104 MPSoC power consumption during inference ranged from 1.5 W to 6.75 W (1.5–1.75 W for lightweight HLS designs and up to 6.75 W for the largest DPU workload), with total board power between 10.5 W and 16 W. These power levels, combined with large latency reductions, yielded energy-per-inference improvements of approximately 11.5$\times$ for the VAE Encoder, 13.91$\times$ for CNetPlusScalar, 7.3$\times$ for ESPERTA, and 2.6$\times$ for LogisticNet. These results indicate that meaningful onboard NN inference can be achieved within typical spacecraft power envelopes.

Overall, the study demonstrated that NN algorithms can be effectively accelerated on FPGAs using Vitis AI and Vitis HLS, achieving lower inference latency and/or reduced power consumption compared to CPU-only execution. Most implementations exceeded real-time spacecraft performance requirements, indicating headroom for further power optimization through frequency scaling, sparse computation, pruning, parameter quantization, or mixed-precision techniques. Future work will focus on deeper parallelization strategies, custom NN accelerator architectures described in HDL, and evaluating the impact of quantization and radiation-induced fault mitigation techniques on performance and reliability.

\section*{ACKNOWLEDGMENT}
This work is supported by the European Commission, with Automatics in Space Exploration (ASAP), project no. 101082633. Generative AI was employed to ensure fluidity in selected text sections. GM would like to acknowledge funding received from the European Union's Horizon 2020 research and innovation programme under the Marie Skłodowska-Curie grant agreement STRIDE 101148539 . GM would also like to  acknowledge funding received from FWO project Helioskill 3E221369.

\bibliographystyle{ieeetr}
\bibliography{biblio}

\end{document}